\documentclass[twoside,fleqn]{article}
\usepackage{espcrc2}

\def\g#1{\gamma_{#1}}
\def\opergx#1#2{(\g{#1}\otimes\xi_{#2})}

\def\g#1{\gamma_{#1}}
\def\operii{(1\otimes1)}
\def\operix#1{(1\otimes\xi_{#1})}
\def\opergi#1{(\g{#1}\otimes1)}
\def\opergx#1#2{(\g{#1}\otimes\xi_{#2})}

\def\gsim{{\mathrel{\raise2pt\hbox to 8pt{\raise -5pt\hbox{$\sim$}\hss{$>$}}}}}
\def\rsim{{\mathrel{\raise2pt\hbox to 8pt{\raise -5pt\hbox{$\sim$}\hss{$>$}}}}}
\def\lsim{{\mathrel{\raise2pt\hbox to 8pt{\raise -5pt\hbox{$\sim$}\hss{$<$}}}}}

\def\etal{{\it et al.}}

\begin{document}

\title{
       \begin{flushright}\normalsize
	    \vskip -0.9 cm
            UW-PT 02-19, SNUTP-02-027
       \end{flushright}
	\vskip -0.4 cm
Matching coefficients for improved staggered bilinears
\thanks{Presented by S.~Sharpe.
Research supported in part by US-DOE contract DE-FG03-96ER40956/A006
and by BK21/KRF-2002-003-C00033.}
}

\author{Weonjong Lee\address{School of Physics,
  Seoul National University,
  Seoul, 151-747, South Korea}
and
Stephen R. Sharpe\address{Physics Department, Box 351560,
University of Washington, Seattle, WA 98195-1560, USA} 
}
      
\begin{abstract}
  We calculate one-loop matching factors
  for bilinear operators composed of improved staggered fermions.
  We compare the results for different improvement schemes used in
  the recent literature, including the HYP action and an action close
  to the Asqtad action.
  We find that all improvement schemes substantially 
  reduce the size of  the one-loop contributions to matching factors.
  The resulting corrections are comparable to, or smaller than,
  those found with Wilson and domain-wall fermions.
\vspace{-0.2in}
\end{abstract}

\maketitle
Staggered fermions provide an attractive option for calculating weak matrix elements.
The main advantages are (i) that
unquenched simulations are significantly faster than with other fermion 
discretizations, and (ii) that the axial symmetry allows the calculation of weak
matrix elements relevant to kaon mixing and decays.
Progress has been slowed, however, by the large $O(a^2)$
errors observed with unimproved staggered fermions---in particular
large staggered-flavor (hereafter ``taste'') 
symmetry breaking---and the large size
of some 1-loop contributions to matching factors.

Both problems are alleviated with improved staggered fermions.
The key ingredient is the use of ``smeared'' links, which reduce the
taste-breaking couplings of high momentum gluons 
to quarks~\cite{Lepageproc,Lagae,Lepage}.
This is one part of the Symanzik improvement program applied at tree 
level~\cite{Lepage}.
Smeared links are expected to substantially reduce the non-perturbative 
taste-symmetry breaking, and this is observed in the pion 
spectrum~\cite{Lagae,OrginosToussaint,Orginos}.
It turns out that the largest reduction is for multiple APE smearing
or HYP links~\cite{HK}.
The smeared links are also expected to reduce the 1-loop contributions
to matching factors, since the taste-changing couplings are responsible
for large tadpole-like contributions~\cite{Golterman}.
The purpose of our present work is to test the latter expectation
with a variety of smearing choices.

We carry out this test using ``hypercube bilinears'', i.e. bilinear operators
in which the quark and antiquark fields reside on the same $2^4$ hypercube.
These are the most local bilinears which match onto the complete set of
continuum operators, having all spins and tastes. They are also the building
blocks for the four-fermion operators we use in calculations of weak matrix elements.

We compare the following choices of improved actions and operators
(and use the associated numbers to refer to them below):\footnote{%
For details of the definitions of the various smeared links we refer
to the original references.}
\begin{enumerate}
\item
Unimproved staggered action
and operators made gauge invariant using the original links.
\item
Unimproved action and operators except that all links are replaced by
the Fat-7 smeared links introduced in Ref.~\cite{OrginosToussaint}.
\item
Unimproved action and operators except that all links are replaced by
the $O(a^2)$ improved links introduced in Ref.~\cite{Lepage}
(these involve Fat-7 smearing and the additional double-staple ``Lepage'' term).
\item
Unimproved action and operators except that all links are replaced by
HYP links~\cite{HK}. We consider two choices of the HYP smearing parameters,
denoted 4A and 4B below.
\item
Asqtad-like action 
(includes the Naik term but uses unimproved Wilson glue instead of the
improved gluon action used by the MILC collaboration~\cite{Orginos})
and operators containing $O(a^2)$ improved links.
\end{enumerate}
We tadpole-improve all links (using the fourth-root of the plaquette to define
$u_0$) except for the HYP links, for which we discuss mean-field
improvement below.

The new feature introduced into perturbative calculations by smearing
is that a link in, say, the $\mu$-th direction couples to gluons
polarized in all directions $\nu$, not only to $\nu=\mu$ as for the original links.
This introduces a vertex factor
\begin{equation}
\delta_{\nu,\mu} D_\mu(k) + (1 - \delta_{\nu,\mu}) G_{\nu,\mu}(k)
\end{equation}
where
\begin{eqnarray}
D_\mu(k) &\!=\!& 1 - d_1 \sum_{\nu\ne\mu} {\bar s}_\nu^2
+ d_2 \sum_{\nu < \rho \atop \nu,\rho\ne\mu}{\bar s}_\nu^2 {\bar s}_\rho^2
\nonumber\\
&&\mbox{}- d_3 {\bar s}_\nu^2 {\bar s}_\rho^2 {\bar s}_\sigma^2 
- d_4 \sum_{\nu\ne\mu} {\bar s}_\nu^4 
\,, 
\end{eqnarray}
with ${\bar s}_\nu= \sin(k_\nu/2)$, etc., and
\begin{eqnarray}
G_{\nu,\mu}(k) &\!=\!& {\bar s}_\mu {\bar s}_\nu \widetilde G_{\nu,\mu}(k) 
\\
\widetilde G_{\nu,\mu}(k) &\!=\!& d_1 
\!-\! d_2 \frac{({\bar s}_\rho^2\!+\! {\bar s}_\sigma^2)}{2}
\!+\! d_3 \frac{{\bar s}_\rho^2 {\bar s}_\sigma^2}{3}
\!+\! d_4 {\bar s}_\nu^2
\,,
\end{eqnarray}
with all indices ($\mu,\nu,\rho,\sigma$) different.
This general form encompasses all our smearing choices and agrees
with previous results~\cite{Hein,HK2}.

The different smearing choices are distinguished by
the values of the coefficients $d_{1-4}$:
\\ \noindent
1. Unimproved:
$ d_1 \!=\! 
d_2 \!=\! 
d_3 \!=\! 
d_4 \!=\! 0
$;
\\ \noindent
2. Fat-7 links:
$
d_1 \!=\! 
d_2 \!=\! 
d_3 \!=\! 1, 
d_4 \!=\! 0
$;
\\ \noindent
3 \& 5.
$O(a^2)$ improved:
$
d_1 \!=\! 0, 
d_2 \!=\! 
d_3 \!=\! 
d_4 \!=\! 1
$;
\\ \noindent
4. HYP links:
We consider two choices for the smearing parameters
which define the HYP link.
\\ \noindent
4A.
Those from Ref.~\cite{HK}, which were 
determined using a non-perturbative optimization procedure:
$
d_1 \!=\! 0.89,
d_2 \!=\! 0.96,
d_3 \!=\! 1.08,
d_4 \!=\! 0
$
\\ \noindent
4B.
The second choice
leads to the {\em same} coefficients as for Fat-7 links
(case 2 above)
and thus removes $O(a^2)$ taste-symmetry breaking.
Note, however, that the two-gluon vertices are not
the same as for the Fat-7 links, so that 
one-loop tadpole diagrams differ.

Using these new vertices, we calculate the one-loop matching 
to continuum bilinears regularized in the NDR scheme:
\begin{equation}
{\cal O}_i^{\rm cont} \!=\! {\cal O}_i^{\rm lat} \!+\!
{C_F g^2\over 16\pi^2} \sum_j
(\delta_{ij} 2 d_i \ln(\mu a) \!+\! c_{ij}) {\cal O}_j^{\rm lat}
\end{equation}
where $d_i= (3,0,-1)$ for spins $(S/P,V/A,T)$,
and the labels $i,j$ run over spin-tastes.
For example, the continuum operators are
\begin{equation}
{\cal O}^{\rm cont}_{i=\opergx{S}{F}} =
\overline Q^1_{\alpha, a} \gamma_S^{\alpha\beta} \xi_F^{ab} Q^2_{\beta, b} \,,\quad
\xi_F=\gamma_F^*\,.
\end{equation}
The superscripts on $Q$ indicate the continuum flavor---we consider only
flavor non-singlets.

The details of the calculation are presented in Ref.~\cite{LeeSharpe}.
To check our results, we do two independent 
calculations using different methods.
We focus here on the diagonal matching coefficients $c_{ii}$---four
of the off-diagonal coefficients are non-vanishing but all are small,
irrespective of smearing.
Results for the $c_{ii}$ are given in Table~\ref{tab:cii}. 
These are the matching coefficients if the NDR scale is taken to be $\mu=1/a$.
This is expected to be within a factor of 2 of the optimum value (``$q^*$''), 
and in any case the coefficients depend only weakly on this scale.
It is sufficient here to quote only one decimal place---more 
accurate results can be found in Ref.~\cite{LeeSharpe}.

\begin{table}
\setlength{\tabcolsep}{1.5mm}
\begin{center}
{\small
\begin{tabular}{crrrrrr}
Operator-i		      	&
	(1)\ &(2)\  &(3)\  &(4A)\  &(4B)\ &(5)\ \\ 
\hline 
$\operii$			&-29.4& 1.9&-4.4&-0.6&-0.1&-2.2\\
$\operix \mu$			& -8.6& 2.5&-2.6& 1.8& 2.5&-0.3\\
$\operix{\mu\nu}$		&  0.6& 2.9&-2.8& 4.0& 4.9&-0.8\\
$\operix{\mu5}$			&  5.2& 3.3&-4.0& 6.0& 7.3&-2.1\\
$\operix{5}$			&  8.7& 3.8&-5.6& 8.0& 9.7&-3.8\\
$\opergi{\mu}$			&  0  & 0  & 0  & 0  & 0  & 1.4\\
$\opergx\mu\mu$			& -4.9& 0.8& 2.9&-0.9&-1.2& 4.3\\
$\opergx\mu\nu$			&  0.2&-0.1&-3.0& 1.3& 1.8&-1.5\\
$\opergx\mu{\mu\nu}$		& -3.4& 0.4&-0.1& 0.3& 0.4& 1.4\\
$\opergx\mu{\nu\rho}$		&  2.5&-0.2&-5.5& 2.7& 3.7&-4.0\\
$\opergx\mu{\nu5}$		&  0.2& 0.1&-2.5& 1.6& 2.1&-1.0\\
$\opergx\mu{\mu5}$		&  4.9&-0.2&-7.9& 4.2& 5.7&-6.5\\
$\opergx\mu{5}$			&  2.8& 0.0&-5.0& 3.0& 4.0&-3.6\\
$\opergi{\mu\nu}$		&  1.6& 0.4&-1.3& 1.9& 2.3&-0.0\\
$\opergx{\mu\nu}\mu$		&  0.8& 0.9& 2.1& 1.0& 0.9& 3.3\\
$\opergx{\mu\nu}\rho$		&  3.0& 0.0&-4.5& 3.0& 4.0&-3.2\\
$\opergx{\mu\nu}{\mu\nu}$	&  4.6& 1.8& 6.7& 0.3&-0.2& 7.8\\
$\opergx{\mu\nu}{\mu\rho}$	&  1.3& 0.4&-1.4& 1.9& 2.3&-0.1\\
$\opergx{\mu\nu}{\rho\sigma}$	&  4.9&-0.2&-7.4& 4.2& 5.7&-6.1\\
\end{tabular}
}
\end{center}
\caption{Diagonal matching coefficients $c_{ii}$ for the improvement
choices listed in the text.
The components $\mu$, $\nu$, $\rho$, $\sigma$ are all different.
\label{tab:cii}}
\vspace{-0.3in}
\end{table}

We see from the table that all the choices of smeared link that we consider
significantly reduce the size of the largest matching coefficients.
Note that, in present simulations with $1/a\approx 2\;$GeV, 
$\alpha_{\overline{\rm MS}}\approx 0.19$ so $c_{ii}=5$ corresponds to 
a $10\%$ correction. Thus the matching coefficient for
the unimproved staggered fermion scalar bilinear ($c=-29.4$) leads to $\sim 60\%$
corrections, so that perturbation theory is not reliable.
For all smearing choices $|c|<10$, which is a considerable improvement,
and makes the corrections comparable to those for Wilson
($c_S\!=\!-0.1$, $\!c_P\!=\!-9.7$, $\!c_V\!=\!-7.8$, 
$\!c_A\!=\!-2.9$, $\!c_T\!=\!-4.3$ in the tadpole improvement scheme
of Ref.~\cite{BGS}) 
and domain-wall fermions
($c_{S/P}\!=\!-11.2$, $c_{V/A}\!=\!-5.3$, $c_T\!=\!-2.0$, with the
domain-wall mass parameter $M=1.7$~\cite{Aoki}).
The ``best'' choice is case 2, i.e. Fat-7 links.

The corrections for HYP fermions can be further reduced using
mean-field improvement. This is essentially a copy of the tadpole
improvement scheme applied to the smeared links. We define
$u_0^{\rm SM}$ to be the fourth-root of the plaquette
built out of 
smeared links. Each smeared link in the action and operators is then 
divided by $u_0^{\rm SM}$. This is straightforward to implement in simulations
and in perturbation theory.
We expect this to reduce the matching factors
by largely removing the residual fluctuations in the smeared links,
but that the reduction should be by a smaller factor 
than for the original tadpole improvement applied to unimproved, 
Fat-7 or $O(a^2)$
links. This is borne out by our results,
which are quoted in Table~\ref{tab:ciiMFI}.
We have also applied this ``second-level''
of mean-field improvement to the Fat-7 and $O(a^2)$ improved links,
for which the effect is to slightly increase the corrections.\footnote{%
It turns out that, after mean-field improvement, Fat-7 and HYP links with
``Fat-7'' coefficients (case 4B) have identical 1-loop matching factors.}

\begin{table}
\begin{center}
{\small
\begin{tabular}{crrr}
Operator-i		      	
	&(2),(4B) &(3)\  &(4A)\  \\ 
\hline 
$\operii$			& 1.0&-8.5&-0.0\\
$\operix \mu$			& 2.5&-2.6& 1.8\\
$\operix{\mu\nu}$		& 3.8& 1.2& 3.4\\
$\operix{\mu5}$			& 5.2& 4.1& 4.9\\
$\operix{5}$			& 6.5& 6.6& 6.2\\
$\opergi{\mu}$			&   0&   0&   0\\
$\opergx\mu\mu$			&-0.1&-1.1&-0.4\\
$\opergx\mu\nu$			& 0.8& 1.1& 0.7\\
$\opergx\mu{\mu\nu}$		& 0.4&-0.1& 0.3\\
$\opergx\mu{\nu\rho}$		& 1.6& 2.6& 1.6\\
$\opergx\mu{\nu5}$		& 1.0& 1.6& 1.0\\
$\opergx\mu{\mu5}$		& 2.5& 4.2& 2.4\\
$\opergx\mu{5}$			& 1.9& 3.1& 1.8\\
$\opergi{\mu\nu}$		& 1.3& 2.8& 1.4\\
$\opergx{\mu\nu}\mu$		& 0.9& 2.1& 1.0\\
$\opergx{\mu\nu}\rho$		& 1.9& 3.6& 1.9\\
$\opergx{\mu\nu}{\mu\nu}$	& 0.8& 2.6& 0.8\\
$\opergx{\mu\nu}{\mu\rho}$	& 1.3& 2.7& 1.4\\
$\opergx{\mu\nu}{\rho\sigma}$	& 2.5& 4.8& 2.5\\
\end{tabular}
}
\end{center}
\caption{Mean-field improved matching coefficients, $c_{ii}^{\rm MF}$
\label{tab:ciiMFI}}
\vspace{-0.3in}
\end{table}

After mean-field improvement, 
the HYP matching factors are nearly as small as those for Fat-7 links
in Table~\ref{tab:cii}. Both have corrections at the 10\% level.
We expect the corrections for
four-fermion operators to be roughly twice this size, and
this is borne out by our preliminary
results for four-fermion operators with HYP links~\cite{LeeLat02}.
Given that HYP links lead to much greater reduction in non-perturbative
taste-symmetry breaking in the pion spectrum, we are encouraged to pursue
our calculations of weak matrix elements using HYP fermions.

\end{document}